\documentclass[12pt]{article}
\usepackage{amsmath,epsfig,amssymb,eufrak}
\usepackage{amscd,latexsym}
\usepackage{color}

\date{}

\def\xlf{\raisebox{+0.2em}{\color{red}\boldmath{$\chi$}}\hspace{-0.2ex}\raisebox{-0.2em}{\color{green}L}
\hspace{-1.5ex}\raisebox{+0.14em}{\color{blue}F}\hspace{2mm}}

\def\lsi{\raise0.3ex\hbox{$<$\kern-0.75em\raise-1.1ex\hbox{$\sim$}}}
\def\gsi{\raise0.3ex\hbox{$>$\kern-0.75em\raise-1.1ex\hbox{$\sim$}}}

\input macros.sty      
\input eqalign.sty     

\begin{document}

\begin{titlepage}

\title{
  {\vspace{-0cm} \normalsize
  \hfill \parbox{40mm}{DESY/05-051\\
                       SFB/CPP-05-10\\March 2005}}\\[10mm]
Light quarks with twisted mass fermions}  
\author{ K.~Jansen$^{\, 1}$, M.~Papinutto$^{\, 1}$, A.~Shindler$^{\, 1}$, \\
C.~Urbach$^{\, 1,2}$ and I.~Wetzorke$^{\, 1}$\\ 
\\
   {\bf \xlf Collaboration}\\
\\
{\small $^{1}$  John von Neumann-Institut f\"ur Computing NIC,} \\
{\small         Platanenallee 6, D-15738 Zeuthen, Germany} \\ \ \\
{\small $^{2}$  Institut f\"{u}r Theoretische Physik, Freie Universit\"{a}t Berlin,} \\
{\small Arnimallee 14, D-14195 Berlin, Germany}
}

\maketitle

\begin{abstract}
We investigate Wilson twisted mass fermions in the quenched approximation
using different definitions of the critical bare quark mass $m_c$ to realize
maximal twist and, correspondingly,
automatic $O(a)$ improvement for physical observables.
A particular definition of $m_c$ is given by extrapolating the value of 
$m_c$ obtained from the PCAC relation at non-vanishing bare twisted quark
mass $\mu$ to $\mu=0$. 
Employing this improved definition of the critical mass
the Wilson twisted mass formulation provides the possibility to perform reliable
simulations down to very small quark masses with correspondingly
small pion masses of $m_\pi \simeq 250 \mathrm{~MeV}$, while keeping
the cutoff effects of $O(a^2)$ under control.
\vspace{0.75cm}
\noindent
\end{abstract}

\end{titlepage}

\section{Introduction}
The combination of an automatic $O(a)$-improvement, the infrared 
regulation of small eigenvalues and fast dynamical simulations render 
the so-called twisted mass fermions \cite{tm,Frezzotti:2003ni,wtmqcd}
a most promising formulation of lattice QCD. 
There are already a number of results for the quenched approximation
\cite{JSUW,ARL,ARLW}. Also full QCD simulations with this 
approach have been performed and revealed a surprising phase 
structure of lattice QCD \cite{tmdyn,dbw2}. On the theoretical side, 
extensive calculations in chiral perturbation theory were done 
\cite{SHARPE-SINGLETON,MUNSTER,Luigi}. 

One of the problems that appeared in the discussion of twisted mass fermions
is the exact definition of the critical bare quark mass. In 
\cite{tmoverlap} it was shown that using the critical value of the bare 
quark mass where the pion mass vanishes, leads to a quark mass dependence
of the pion decay constant $f_\pi$ that deviates strongly from the 
expected linear
behavior. This  ''bending phenomenon'' was observed when the quark mass
$m_q$ obeys the inequality
\begin{equation}
am_q < a^2\Lambda^2\; .
\label{inequality}
\end{equation}
The bending phenomenon was interpreted in a way that for such small values
of the quark mass the Wilson term becomes the dominant term in the lattice 
Wilson-Dirac operator, thus leading to large lattice artefacts. 

In refs.~\cite{AokiBaer,SharpeWu} it has been suggested to take
the value of the critical mass where the PCAC quark mass vanishes.
In chiral perturbation theory it was demonstrated that this
definition leads to a stable situation of ''maximal twist'' and hence to a 
reduction of lattice artefacts that otherwise would be enhanced at very small
values of the quark mass. In refs.~\cite{FMPR,Shindler} the theoretical
background of such a definition is discussed on a general ground.

In this paper, we employ the definition of the critical mass by computing 
{\em at non-zero twisted mass} the critical hopping parameter as a functions of
the twisted mass parameter and finally taking the limit to zero twisted mass.
Such a definition is expected to eliminate large $O(a^2)$ artifacts
at small values of the pion mass and should hence avoid the bending phenomenon.
In the following we will shortly describe the encouraging results
employing this improved definition of the critical mass. A more detailed
scaling analysis of mesonic quantities will be reported 
elsewhere \cite{scaling}.

\section{Wilson twisted mass fermions}
In this paper we will work with Wilson twisted mass fermions that can be 
arranged to be $O(a)$ improved without employing specific improvement terms
\cite{tm,Frezzotti:2003ni}. The Wilson tmQCD action in the twisted basis 
can be written as
\begin{equation}
  \label{tmaction}
  S[U,\psi,\bar\psi] = a^4 \sum_x \bar\psi(x) ( D_W + m_0 + i \mu
\gamma_5\tau_3 ) \psi(x)\; ,
\end{equation}
where the Wilson-Dirac operator $D_{\rm W}$ is given by
\be
D_{\rm W} = \sum_{\mu=0}^3 \frac{1}{2} 
[ \gamma_\mu(\nabla_\mu^* + \nabla_\mu) - a \nabla_\mu^*\nabla_\mu]
\label{Dw}
\ee
and $\nabla_\mu$ and $\nabla_\mu^*$ denote the usual forward
and backward derivatives. We refer to \cite{tmoverlap} for further
unexplained notations. The definition of the critical mass $m_c$ will be
discussed in detail in the next section.

We extract pseudoscalar and vector meson masses from the correlation
functions at full twist ($m_0=m_c$):
\begin{eqnarray*}
C_P^a(x_0) &=& a^3\sum_{\mathbf x} \langle P^a(x)P^a(0)\rangle \quad a=1,2\\
C_A^a(x_0) &=& \frac{a^3}{3}\sum_{k=1}^3\sum_{\mathbf x} \langle A_k^a(x)A_k^a(0)\rangle 
\quad a=1,2 \\
C_T^a(x_0) &=& \frac{a^3}{3}\sum_{k=1}^3\sum_{\mathbf x} \langle T_k^a(x)T_k^a(0)\rangle
\quad a=1,2 
\end{eqnarray*}
where we consider the usual local bilinears 
$P^a=\bar{\psi}\gamma_5\frac{\tau^a}{2}\psi$,
$A_k^a=\bar{\psi}\gamma_k\gamma_5\frac{\tau^a}{2}\psi$ and
$T_k^a=\bar{\psi}\sigma_{0 k}\frac{\tau^a}{2}\psi$.

The untwisted PCAC quark mass $m_{\textrm{PCAC}}$ can be extracted from the ratio
\be
m_{\textrm{PCAC}}=\frac{\sum_{\mathbf x}\langle
\partial_0 A_0^a(x)\; P^a(0)\rangle}{2\sum_{\mathbf x}\langle 
P^a(x)P^a(0)\rangle}\quad a=1,2\; .
\label{mPCAC}
\ee

Using the PCAC relation we can also compute the pion decay constant
at maximal twist without requiring any renormalization constant 
(see \cite{ALPHA,JSUW}):
\be
f_\pi=\frac{2\mu}{(M_\pi)^2} | \langle
0 | P^a | \pi\rangle | \qquad a=1,2 \; .
\label{indirect}
\ee

\section{Definition of the critical mass}
The Wilson tmQCD action of eq.~(\ref{tmaction}) can be studied in the full
parameter space $(m_0,\mu)$. A special case arises, however, when
$m_0$ is tuned towards a critical bare quark mass $m_{\mathrm{c}}$.
In such, and only in such a situation, all physical quantities are, or
can easily be, $O(a)$ improved. The critical quark mass, or 
alternatively the critical hopping parameter 
$\kappa_c = \frac{1}{2 a m_c+8}$, has thus to be fixed in the actual simulation
to achieve automatic $O(a)$-improvement at maximal twist.

In general, the definition of the critical mass has an intrinsic uncertainty
that comes at $O(a)$ for Wilson fermions. This could be improved to an
$O(a^2)$ uncertainty by using Clover fermions.
If a definition of $\kappa_c$ from a linear extrapolation of 
$(m_\pi a)^2 \to 0$  is used, one has in addition an unrelated systematic error
coming from the long way of extrapolation from large pion masses of O(600 MeV).
\\

\begin{figure}[htb]
\vspace{-0.0cm}
\begin{center}
\epsfig{file=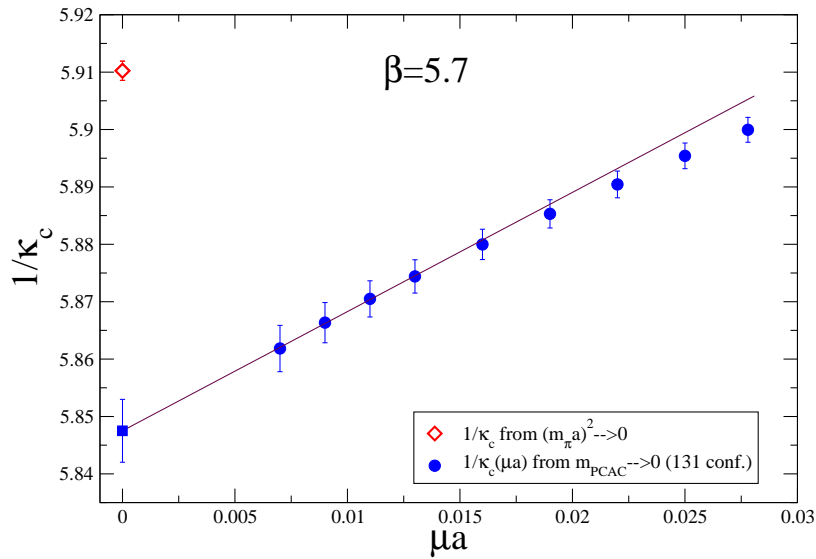,angle=270,width=0.8\linewidth}
\end{center}
\vspace{-0.0cm}
\caption{Determination of the critical mass: $1/\kappa_c$ versus $\mu a$ at 
$\beta=5.70$,
extrapolation to $\mu a = 0$, the open diamond indicates the $1/\kappa_c$
value determined by $(m_\pi a)^2 \to 0$ at $\mu=0$ for unimproved Wilson
fermions.
\label{fig:newkc}}
\end{figure}

A better definition of $\kappa_c$ can be obtained by the following 
procedure:
\begin{itemize}
\item At fixed non-zero twisted mass parameter 
the hopping parameter $\kappa_c(\mu a)$ is determined as the point
where the PCAC quark mass of eq.~(\ref{mPCAC}) vanishes. The non-zero
value of the twisted mass allows a safe interpolation in this case.

\item As a further step, an only short linear extrapolation of 
$\kappa_c(\mu a)$ from small values of $\mu a$ to $\mu a = 0$ yields 
a definition of $\kappa_c$ which is expected \cite{FMPR,Shindler} to lead to
small lattice artefacts, in particular also at small quark masses much below
values indicated in the inequality of eq.~(\ref{inequality}).
\end{itemize}
In figure
\ref{fig:newkc} we show an example of this extrapolation at $\beta=5.7$,
where we also indicate the $\kappa_c$-value determined by the vanishing pion
mass squared.

The possible choices for the determination of the critical 
bare quark mass were also 
discussed in chiral perturbation theory \cite{AokiBaer, SharpeWu}. Recently,
a definition of maximal twist from parity conservation has also been 
investigated in \cite{ARLW}. There, however, the critical masses $m_c(\mu a)$ 
were not extrapolated to $\mu a = 0$, but were used at the respective
twisted mass parameter at which they were determined.

\section{Numerical results}
In this section we will provide a comparison of Wilson twisted mass
results for the pion mass, the pion decay constant and the
$\rho$-mass obtained with two different definitions of $\kappa_c$. 
The first, ``pion mass'' definition, refers to the limit 
$(m_\pi a)^2 \to 0$ taken at $\mu a=0$. 
The second, ``PCAC quark mass'' definition, corresponds to the determination of
$\kappa_c(\mu a)$ at fixed $\mu a \ne 0$ from the vanishing of the PCAC quark
mass, $am_\mathrm{PCAC}(\mu a)$ and then taking the limit 
$\kappa_c(\mu a \rightarrow 0)$. The latter $\kappa_c$ values 
were determined on a subset of $O(150)$ configurations for three couplings,
namely $\beta=5.70, 5.85$ and $6.0$. The simulations were performed for
a number of bare quark masses in a corresponding pion mass range of 
$250 \mathrm{~MeV} < m_\pi < 1200 \mathrm{~MeV}$
using a multiple mass solver \cite{MMS} on $O(400)$ gauge field configurations
generated with the Wilson plaquette gauge action. 

In general the time component of the tensor correlator
provides a better signal for the vector meson mass compared to the axial-vector
correlator at maximal twist. This observation was made independently in
\cite{ARLW} and is manifest in the smaller error bars displayed in figures of
this section.

\begin{figure}[htb]
\vspace{-0.0cm}
\begin{center}
\epsfig{file=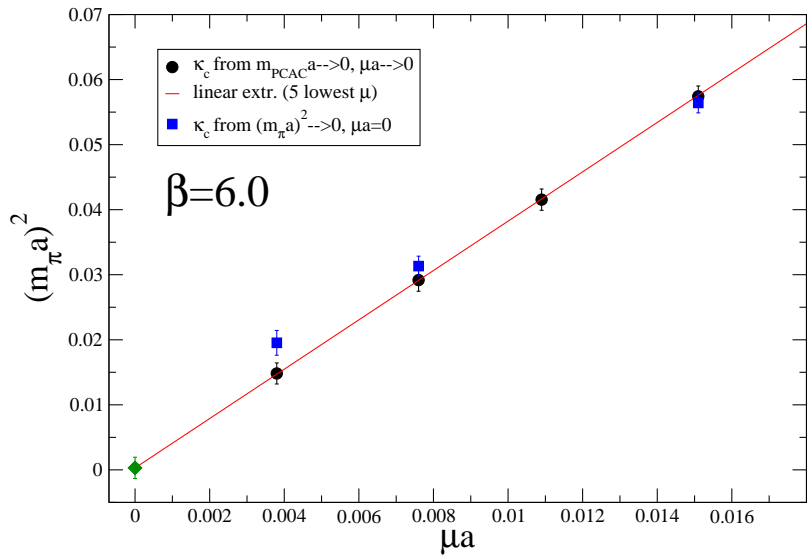,angle=270,width=0.8\linewidth}
\end{center}
\vspace{-0.0cm}
\caption{Comparison of results at $\beta=6.0$ for the pion mass for different 
definitions of the critical mass. \label{fig:pion}}
\end{figure}

The first quantity we investigated was the pion mass. In figure \ref{fig:pion}
we show the behavior of $(m_\pi a)^2$ versus $\mu a$ for the two definitions
of $\kappa_c$. The results for the PCAC quark mass definition 
show a linear behavior down to very small bare
quark masses, while a small non-zero value is approached in the chiral limit
with the pion mass definition of $\kappa_c$. This residual $O(a)$
pion mass at $\mu a = 0$ can be attributed to the $O(a)$ error
in the critical mass determined by the vanishing of the pion mass 
in the pure Wilson case. 

\begin{figure}[htb]
\vspace{-0.0cm}
\begin{center}
\epsfig{file=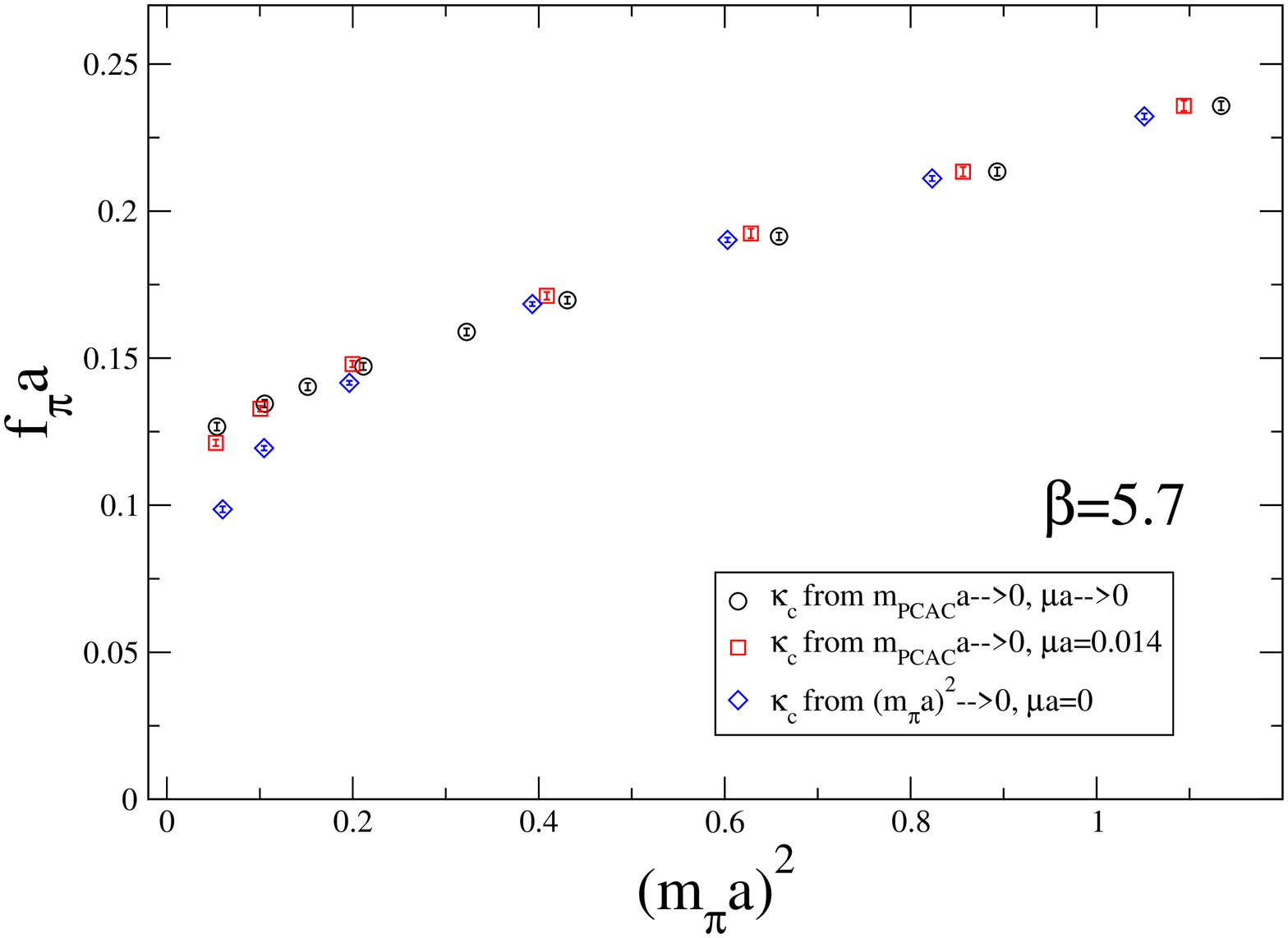,angle=270,width=0.8\linewidth}
\epsfig{file=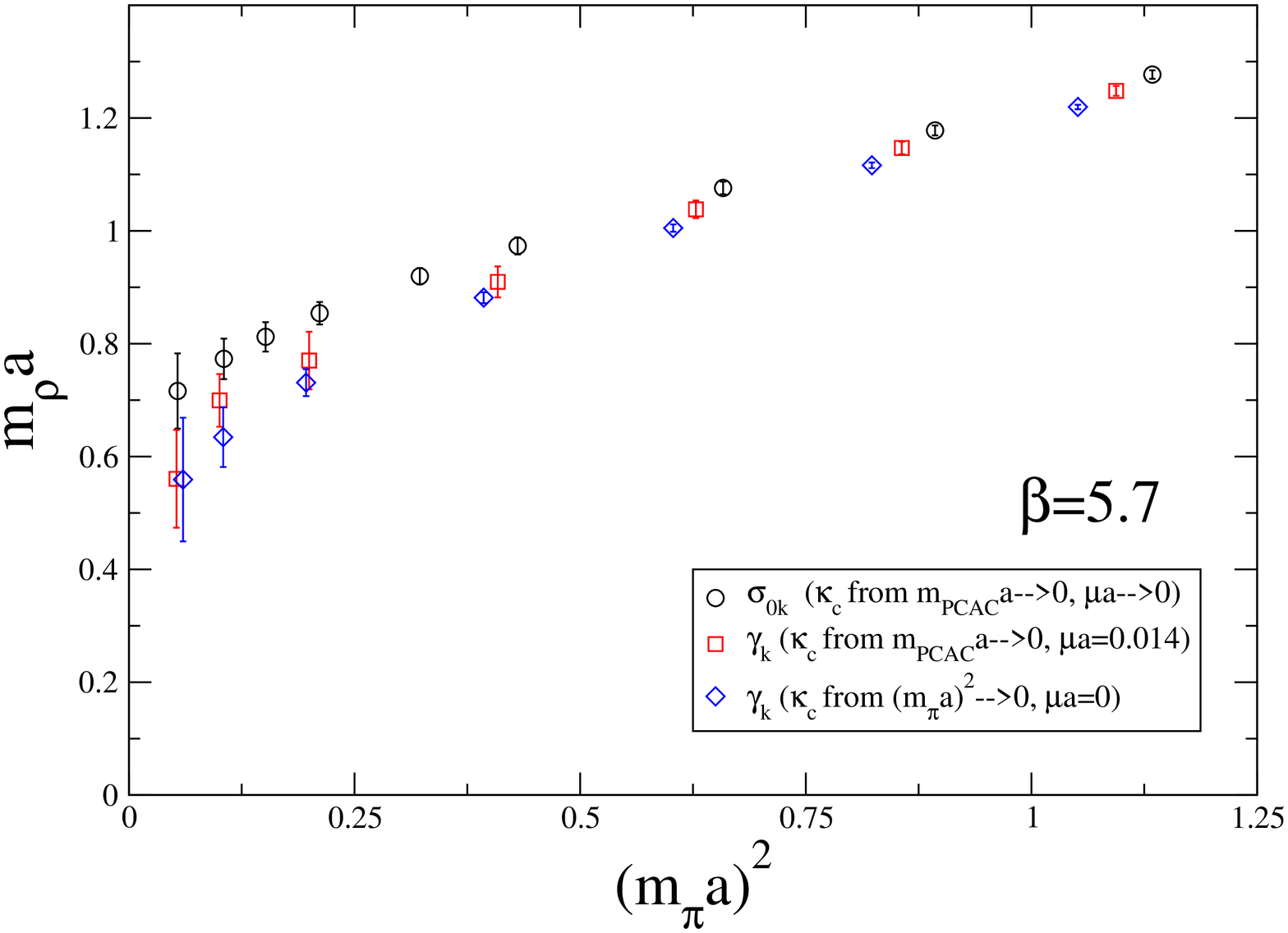,angle=270,width=0.8\linewidth}
\end{center}
\vspace{-0.0cm}
\caption{Comparison of results at $\beta=5.70$ for the pion decay constant (top)
and vector meson mass (bottom) for different definitions of the critical
mass.  \label{fig:compare5.70}}
\end{figure}

At $\beta=5.7$ (lattice size $12^3 \times 32$, $a\approx 0.17 \mathrm{~fm}$) 
we computed correlation functions also for a third, intermediate
definition of the critical mass: we chose the point where the PCAC quark mass 
vanishes at small, but non-zero value of the twisted mass. In figure
\ref{fig:compare5.70} we show the results for the pion decay constant and the
vector meson mass. It is obvious that already this intermediate definition
of $\kappa_c$ substantially diminishes the bending tendency of the data when
approaching the chiral limit. Such a definition of the critical hopping
parameter $\kappa_c(\mu_0 \, a \ne 0)$ seems to be sufficient to eliminate
the deviation from a linear behavior for all twisted mass values 
$\mu a \ge \mu_0 \, a$. The results for $\kappa_c(\mu a = 0)$ 
from the PCAC quark mass definition show a straight behavior for both
observables down to pion masses of about 250 MeV. The bending phenomena
reported in \cite{tmoverlap} can thus be attributed to the 
pion mass definition of the critical mass in this simulation.
Note that at $\beta=5.7$ we are working at a rather coarse value of the
lattice spacing.

\begin{figure}[htb]
\vspace{-0.0cm}
\begin{center}
\epsfig{file=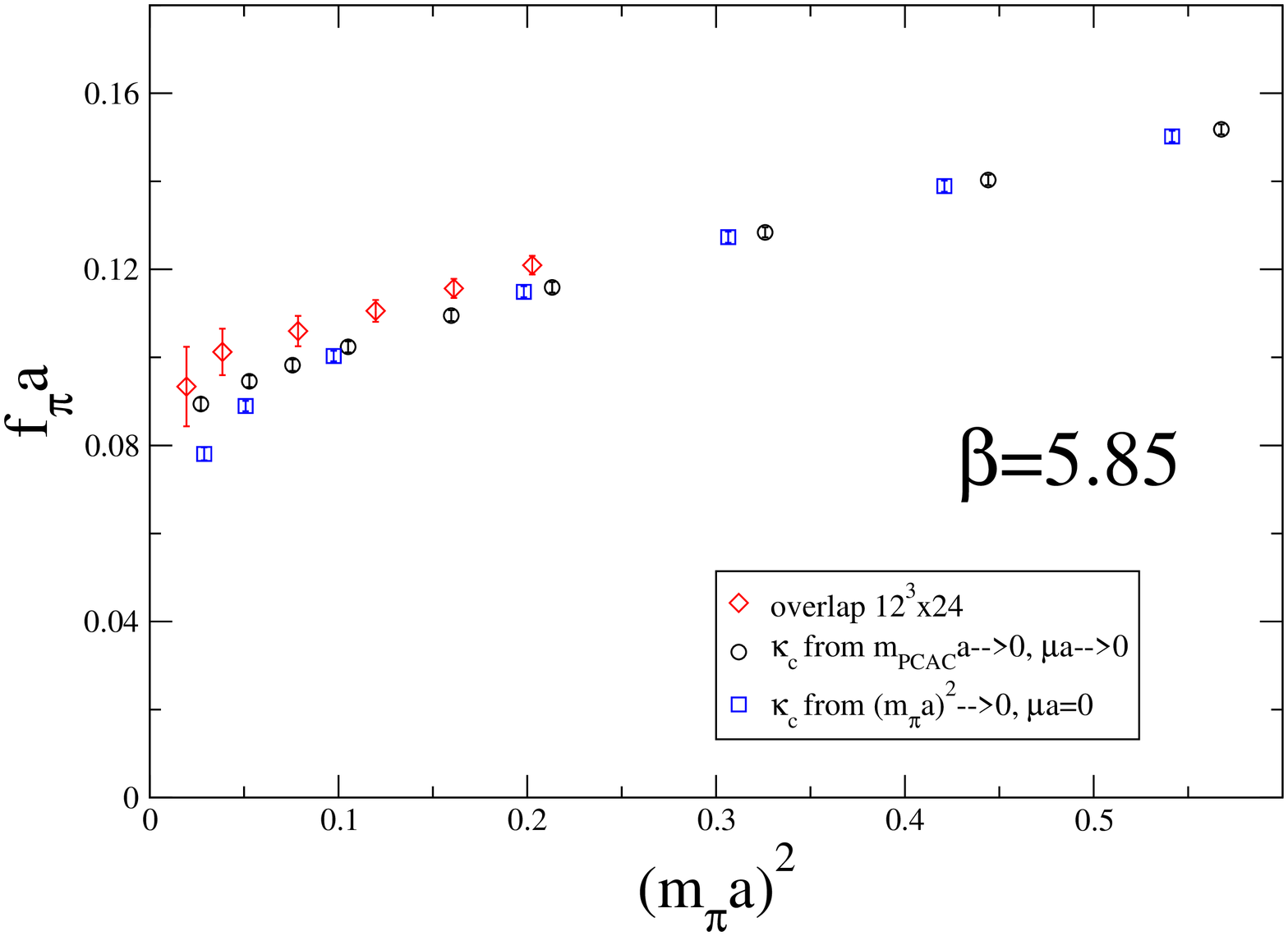,angle=270,width=0.8\linewidth}
\epsfig{file=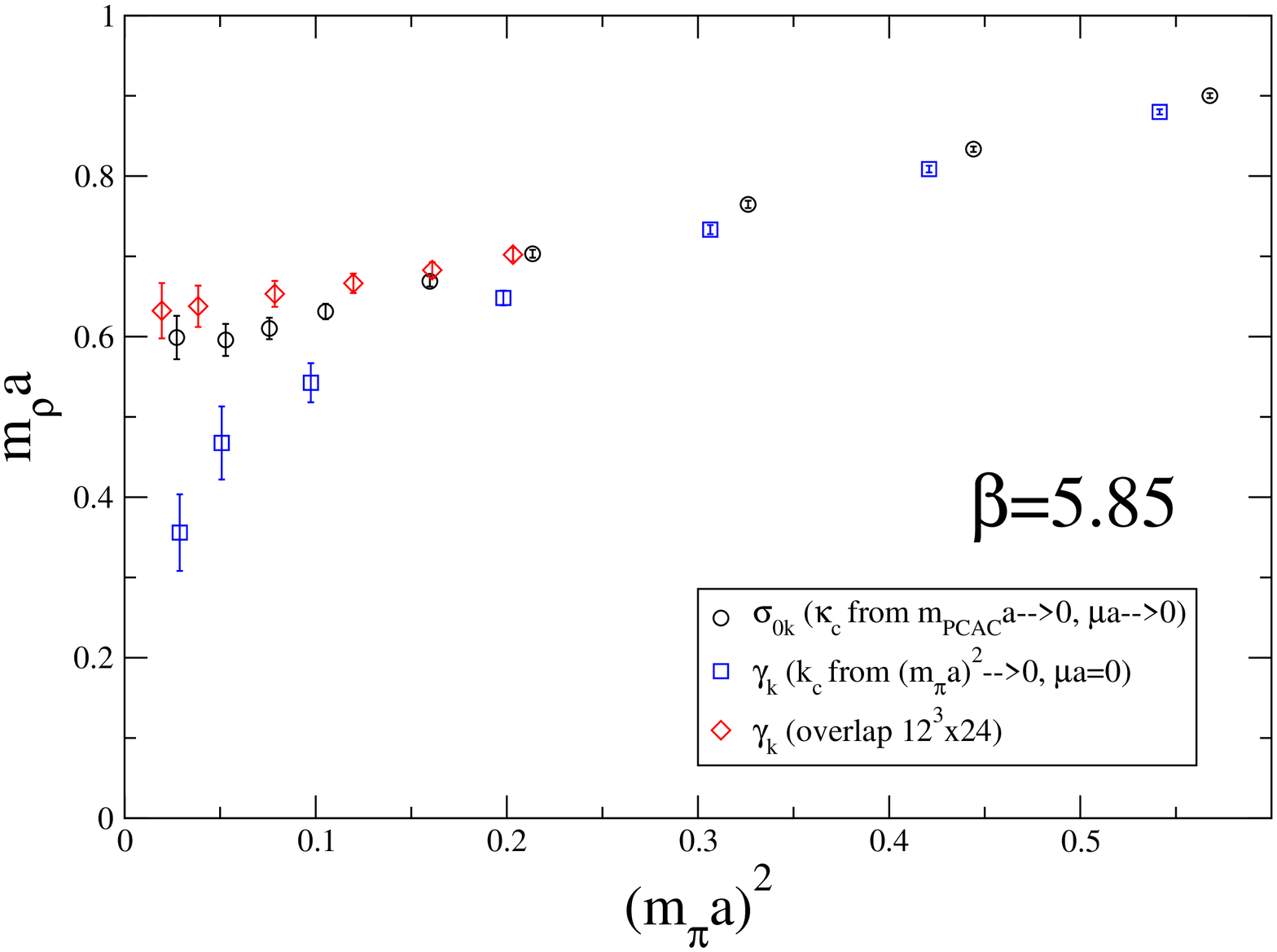,angle=270,width=0.8\linewidth}
\end{center}
\vspace{-0.0cm}
\caption{Comparison of overlap fermion and Wilson twisted mass results at 
$\beta=5.85$ for the pion decay constant (top)
and vector meson mass (bottom) for different definitions of the critical
mass.  \label{fig:compare5.85}}
\end{figure}

At $\beta=5.85$ (lattice size $16^3 \times 32$, $a\approx 0.12 \mathrm{~fm}$) 
we can now compare our previous and new Wilson twisted mass
results with the data from overlap fermion simulations \cite{tmoverlap}
which were obtained on a smaller lattice volume ($12^3 \times 24$).
Using the definition of $\kappa_c$ from the vanishing of the PCAC quark mass,
the bending near the chiral limit vanishes almost completely 
both for the pion decay constant and the vector meson mass
extracted from the tensor correlator (see figure \ref{fig:compare5.85}).
In both cases the overlap fermion results lie slightly higher than the 
twisted mass fermion results. This fact might be explained by 
residual finite volume effects and/or different lattice artefacts.

\begin{figure}[htb]
\vspace{-0.0cm}
\begin{center}
\epsfig{file=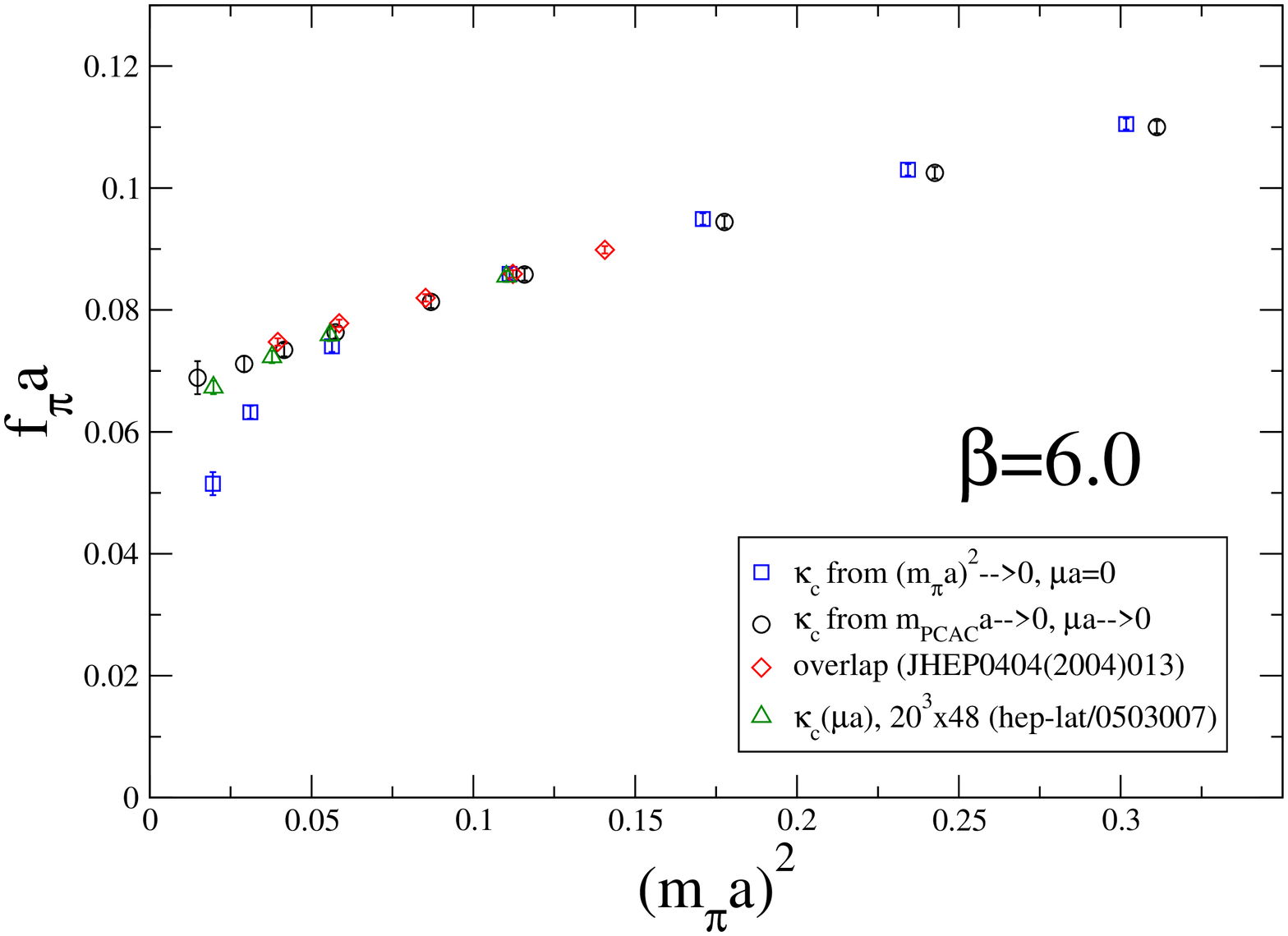,angle=270,width=0.8\linewidth}
\epsfig{file=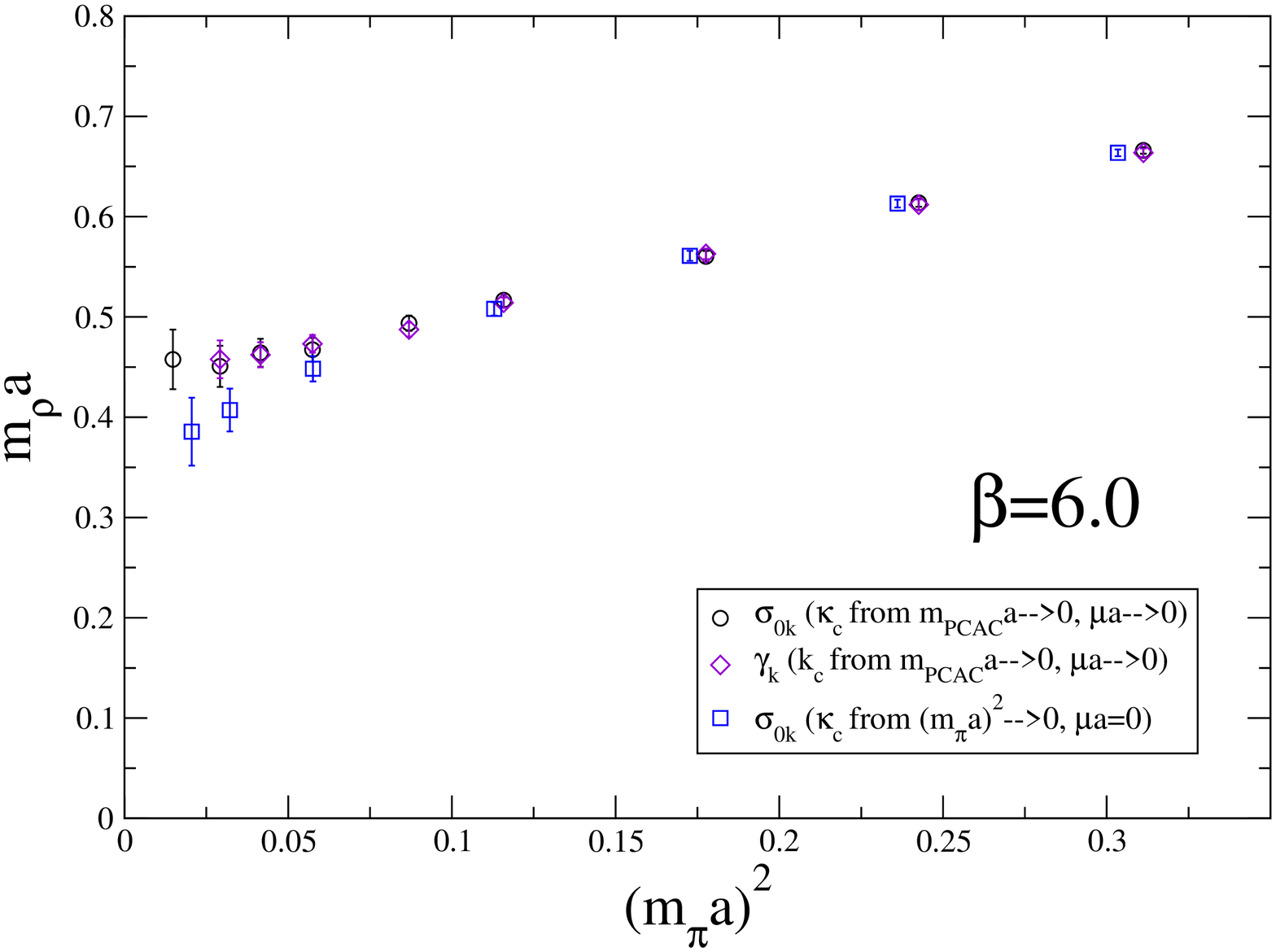,angle=270,width=0.8\linewidth}
\end{center}
\vspace{-0.0cm}
\caption{Comparison of results at $\beta=6.0$ for the pion decay constant 
(top) and vector meson mass (bottom) for different definitions of the critical
mass.  \label{fig:compare6.0}}
\end{figure}

Finally we show in figure \ref{fig:compare6.0} the results at $\beta=6.0$ 
(lattice size $16^3 \times 32$, $a\approx 0.09 \mathrm{~fm}$). 
Although the physical volume is smaller
than the one at $\beta=5.85$ we are able to confirm the absence of bending
for small masses when using the PCAC quark mass definition of $\kappa_c$. 
For the pion decay constant we observe a very good agreement with the
overlap fermion results of ref.~\cite{Giusti} and with the twisted mass
results using the parity conservation definition of the critical mass
from ref.~\cite{ARLW}. For the vector meson mass we refrain from comparing
with the parity conservation definition, since the error bars would cover
both our results with the pion mass and PCAC quark mass definition of the
critical mass. 

\section{Conclusions}
We have studied Wilson twisted mass fermions with different definitions
of the critical mass.
The ''bending phenomena'' in basic observables like the vector meson mass and 
the pion decay constant using the critical mass from vanishing pion mass at 
$\mu a=0$ reported in \cite{tmoverlap} is absent for the
choice of $\kappa_c$ from the vanishing of the PCAC quark mass in the 
limit $\mu a \to 0$. This is true even at small values of the pion mass close 
to the experimentally measured value where maybe contact to chiral perturbation
theory can be made. 

In ref.~\cite{ARLW}, the same definition of the critical mass has been
employed, {\em but at fixed twisted mass parameter separately at each
simulation point}. The results of this reference taken together with the
findings in the present paper lead to the conclusion 
that the PCAC quark mass definition of the critical 
hopping parameter is a crucial element of twisted mass simulations
in order to keep $O(a^2)$ effects well under control at small
quark masses. This conclusion is in accordance with theoretical
considerations \cite{AokiBaer, SharpeWu, FMPR, Shindler}.

Employing the PCAC quark mass definition of the critical mass the Wilson
twisted mass formulation provides the possibility to perform reliable
simulations at very small quark masses ($m_\pi \simeq 250 \mathrm{~MeV}$). 
Thus, Wilson twisted mass fermions can be used to explore really light quarks 
on the lattice - as light as with overlap fermions, but at considerably lower 
cost (see \cite{solver} for a detailed cost comparison). 
This opens a very promising
prospect for dynamical fermion simulations and renders twisted mass fermions as
a real alternative to staggered fermions without the conceptual difficulties of
the latter.

\section{Acknowledgments}
We thank R.~Frezzotti, G.~C.~Rossi, L.~Scorzato, S.~Sharpe and U.~Wenger
for many useful discussions. 
The computer centers at NIC/DESY Zeuthen, NIC at Forschungszentrum
J{\"u}lich and HLRN provided the necessary technical help and computer
resources. This work was supported by the DFG 
Sonderforschungsbereich/Transregio SFB/TR9-03.

\end{document}